\documentclass[twocolumn,preprintnumbers,amssymb,amsmath,aps,floatfix,prc,nofootinbib,superscriptaddress]{revtex4}

\usepackage{epsfig}
\usepackage{bm}
\usepackage{amssymb}
\usepackage{amsmath}
\usepackage{color}
\usepackage{subfigure}
\usepackage{hyperref}

\begin{document}

\title{(3+1)-D viscous hydrodynamics CLVisc at finite net baryon density: identified particle spectra, anisotropic flows and flow fluctuations across BES energies}

\author{Xiang-Yu Wu}
\email{xiangyuwu@mails.ccnu.edu.cn}
\affiliation{Institute of Particle Physics and Key Laboratory of Quark and Lepton Physics (MOE), Central China Normal University, Wuhan, 430079, China}

\author{Guang-You Qin}
\email{guangyou.qin@mail.ccnu.edu.cn}
\affiliation{Institute of Particle Physics and Key Laboratory of Quark and Lepton Physics (MOE), Central China Normal University, Wuhan, 430079, China}

\author{Long-Gang Pang}
\email{lgpang@mail.ccnu.edu.cn}
\affiliation{Institute of Particle Physics and Key Laboratory of Quark and Lepton Physics (MOE), Central China Normal University, Wuhan, 430079, China}

\author{Xin-Nian Wang}
\email{xnwang@lbl.gov}
\thanks{Current address: 2.}
\affiliation{Institute of Particle Physics and Key Laboratory of Quark and Lepton Physics (MOE), Central China Normal University, Wuhan, 430079, China}
\affiliation{Nuclear Science Division, Lawrence Berkeley National Laboratory, Berkeley, CA 94720, USA}

\begin{abstract}

To study the bulk properties of the quark-gluon-plasma (QGP) produced at the beam energy scan (BES) energies at the Relativistic Heavy Ion Collider (RHIC), we extend the (3+1)-dimensional viscous hydrodynamics CLVisc to include net baryon number conservation and Israel-Stewart-like equation for baryon diffusion with the NEOS-BQS equation of state, fluctuating initial conditions from Monte-Carlo Glauber model, and the afterburner SMASH.
This integrated framework is shown to provide a good description of identified particle spectra, mean transverse momenta and anisotropic flows for different centralities and over a wide range of collision energies (7.7-62.4~GeV).
It is found that the mean momenta of identified particles and anisotropic flows increases mildly with the collision energy due to larger radial flow.
We further compute the multiple-particle cumulant ratio $v_2\{4\}/v_2\{2\}$ of elliptic flow across BES energies, and find that the relative fluctuations of elliptic flow are insensitive to the collision energy, consistent with the preliminary STAR data.
Our model provides a benchmark for understanding the RHIC-BES data and studying the critical properties and phase structure of hot and dense QCD matter.

\end{abstract}
\maketitle

\section{Introduction}

Quark Gluon Plasma (QGP), a novel state of matter consisting of deconfined quarks and gluons, can be created under extreme temperature and density environments, such as colliding heavy nuclei at ultra-relativistic energies at the Large Hadron Collider (LHC) and the Relativistic Heavy-Ion Collider (RHIC).
The exploration of such hot and dense QCD matter is one of the most important objectives of high-energy nuclear physics.
Numerical simulations from lattice QCD have demonstrated that the transition from QGP to hadronic phase is a rapid smooth crossover at vanishing baryon density and high temperature \citep{Borsanyi:2013bia, Bazavov:2014pvz}.
In baryon-rich regions at low collision energies, lattice QCD, however, is not effective due to the famous sign problem.
Various theoretical models have predicted a first order transition at large values of the baryon chemical potential \citep{Wu:2021xgu, Fukushima:2010bq}.
This indicates the existence of a critical point between the crossover and the first-order phase transition in  the phase diagram.
In fact, one of the main purposes of beam energy scan (BES) experiments performed at RHIC is to locate the critical point and explore the phase diagram of QCD \citep{Stephanov:2004wx,Stephanov:2006zvm,Stephanov:2011pb,Stephanov:2008qz}.

Experiments have observed many important evidences for the formation of QGP, one of which is strong collective anisotropic flows which lead to azimuthal asymmetry in the final-state hadron momentum distribution.
Relativistic hydrodynamics has been very successful in describing the space-time evolution of QGP and in explaining the observed anisotropic collective flow phenomena \citep{Romatschke:2017ejr, Rischke:1995ir, Heinz:2013th, Gale:2013da, Huovinen:2013wma,Pang:2012he,Wu:2018cpc,Zhao:2017yhj,Pang:2018zzo,Ding:2021ajz,Ze-Fang:2017ppe,Shen:2014vra}.
Due to the strong interaction among QGP constituents, the initial geometric anisotropies of the fireball are translated into final state momentum anisotropies of soft hadrons \cite{Gyulassy:1996br, Aguiar:2001ac, Broniowski:2007ft, Andrade:2008xh, Hirano:2009ah, Alver:2010gr, Petersen:2010cw, Qin:2010pf, Staig:2010pn, Teaney:2010vd}.
One of the main findings from heavy-ion experiments at RHIC and the LHC is that the produced QGP behaves like a nearly-perfect fluid with extremely low ratio of shear viscosity and entropy density \citep{Adams:2003zg, Aamodt:2010pa, ATLAS:2011ah, Chatrchyan:2012ta}.

The framework of relativistic viscous hydrodynamics has provided a powerful tool to study the bulk observables, such as high-order anisotropic flows, flow fluctuations \citep{Giacalone:2017uqx,Rao:2019vgy}, flow correlations \citep{ALICE:2016kpq, Zhu:2016puf, Giacalone:2016afq,Yan:2015jma, Qian:2016fpi, Acharya:2017zfg}, and even collective flows in small collision systems \citep{Zhao:2017rgg, Qin:2013bha, Schenke:2014zha, Bzdak:2014dia, Zhao:2019ehg,Schenke:2021mxx,Bozek:2011if,Bozek:2013uha,Bozek:2016kpf,Nagle:2018nvi}, at RHIC top energies and at the LHC energies, where the net baryon density is nearly vanishing.
However, in heavy-ion collisions at the BES low energies, the assumption of vanishing baryon chemical potential is not valid any more.
The evolution of conserved baryon current needs to be included, since the baryon diffusion has strong effects on the spectra, elliptic flow and rapidity distribution of baryons and anti-baryons \cite{Denicol:2018wdp, Wu:2021ypv}.
In addition, two nuclei have finite overlap time to interact with each other at low collision energies.
This means that the pre-equilibrium stage \citep{Denicol:2018wdp,Kurkela:2018vqr,Kurkela:2018wud} and the longitudinal dynamics are more important for understanding the evolution of the full collision system \citep{Shen:2017bsr,Karpenko:2015xea,Akamatsu:2018olk}.
On the other hand, the equation of state (EoS) has to take into account the finite baryon density in order to consistently close the equations of motion of the fluid.
During the last few years, several models have been developed to take into account these issues and perform realistic hydrodynamics simulation at finite baryon density \citep{Shen:2020mgh,Akamatsu:2018olk,Cimerman:2020iny,Wu:2021ypv,Abdel-Waged:2020kfa,Du:2019obx,Shen:2020jwv}.

In this paper, we present a new integrated event-by-event framework by extending the (3+1)-dimensional CLVisc viscous hydrodynamics to include finite net baryon density with the NEOS-BQS equation of state, fluctuating initial conditions from Monte-Carlo Glauber model and the afterburner SMASH \citep{Weil:2016zrk}.
CLVisc is the most advanced event-by-event viscous hydrodynamics with parallelization of both hydrodynamic equation solver and Cooper-Frye particilization on graphics processing units (GPUs) using OpenCL languages \cite{Pang:2018zzo}.
It has been widely used in the literature, such as the recent work by JETSCAPE \cite{JETSCAPE:2021ehl}.
Due to the acceleration of GPUs, CLVisc is one of the fastest viscous hydrodynamics codes in the field.
This makes it possible to develop the CoLBT-hydro model which can simulate the evolution of jets and QGP and their interaction cocurrently, see e.g., \cite{Yang:2021qtl}.
The extension of CLVisc to finite net baryon number density region is an important step forward for studying a lot of important physics at BES energies, such as collectivity, vorticity and polarization, jet-medium interaction, etc.
In this work, we apply our integrated framework to study the collectivity of the QGP produced in heavy-ion collisions at RHIC-BES energies.
The centrality and beam energy dependences of identified particle spectra, mean transverse momenta, anisotropic flows and the relative fluctuations of elliptic flow are studied in detail.

The paper is organized as follows. In Sec.\ref{model}, we extend the event-by-event (3+1)-dimensional viscous hydrodynamics model CLVisc \citep{Pang:2018zzo} to include the evolution equations of net baryon conservation and dissipative baryon current.
The setup for the 3-dimensional fluctuating Monte-Carlo Glauber initial condition model is presented.
The NEOS-BQS equation of state \citep{Monnai:2019hkn,Monnai:2021kgu}, the freeze-out and particlization, and the SMASH \citep{Weil:2016zrk,Schafer:2019edr,Mohs:2019iee,Hammelmann:2019vwd,Mohs:2020awg} afterburner are also briefly introduced.
In Sec. \ref{results}, we present our numerical results for the centrality and collision energy dependences of identified particle spectra, anisotropic flows and flow fluctuations in Au+Au collision at BES energies (7.7-62.4~GeV).
Sec. \ref{summary} contains our summary.

\section{Event-by-event (3+1)-dimensional hydrodynamics CLVisc at finite net baryon density}
\label{model}

In this section, we will describe various components in our integrated event-by-event (3+1)-dimensional hydrodynamics model CLVisc at finite net baryon density, including the Monte-Carlo Glauber initial conditions, the viscous hydrodynamic evolution, the  equation of state, the freeze-out and the hadronic transport model SMASH.

\subsection{Initial condition}

In this work, we use the Monte-Carlo Glauber model to provide initial conditions for the hydrodynamic evolution.
The local entropy density $s(x,y,\eta_s)$ and local baryon density $n(x,y,\eta_s)$ at the initial proper time $\tau_0$ are obtained according to the following expressions \citep{Denicol:2018wdp}:
\begin{eqnarray}
s(x,y,\eta_s) = \frac{K}{\tau_0}(H^{s}_P(\eta_s)s_P(x,y)+H^{s}_T(\eta_s)s_T(x,y))\,,   \\
n(x,y,\eta_s) = \frac{1}{\tau_0}(H^{n}_P(\eta_s)s_P(x,y)+H^{n}_T(\eta_s)s_T(x,y))\,,
\end{eqnarray}
where $s_P(x,y)$ and $s_T(x,y)$ are the entropy densities in the transverse plane produced from the projectile and target nuclei,
$H_P^s(\eta_s)$, $H_T^s(\eta_s)$, $H_P^n(\eta_s)$ and $H_T^n(\eta_s)$ are longitudinal envelop functions which describe the rapidity dependences for the initial entropy density and baryon number density associated with  projectile and target nuclei, $K$ is the scale factor to control the magnitude of initial entropy, and $\tau_0$ is the initial proper time for the hydrodynamic evolution.
Since the Lorentz contraction is smaller at low collision energies, the time for two nuclei to pass each other is larger.

One may estimate the overlap time as $\tau_{\rm overlap} = \frac{2R}{\sinh(y_{\rm beam})}$ , where $R$ denotes the radius of the nucleus and $y_{\rm beam}$ represents the beam rapidity.
{  In this work, we do not include the pre-equilibrium stage, which becomes more important in lower collision energies~\citep{Dore:2020fiq}. To compensate some effects from the pre-equilibrium evolution, we choose the initial proper time $\tau_0$ of hydrodynamics evolution to be larger than the overlap time $\tau_{\rm overlap}$ of two nuclei so that the system has more time to approach the local equilibrium before hydrodynamics starts.}

As for the entropy density $s_{P/T}(x,y)$ in the transverse plane, it is the sum of the Gaussian smearing functions over all the participating  projectile and target nucleons in the Monte-Carlo Glauber model:
\begin{equation}
s_{P/T}(x,y) = \sum_{i}^{P/T} \frac{1}{2\pi\sigma_r^2}\exp\frac{-(x_i-x)^2-(y_i-y)^2}{2\sigma_r^2} \, ,
\end{equation}
where $(x_i, y_i)$ is the transverse position of the participant nucleon and $\sigma_r$ is the transverse Gaussian smearing width (we take $\sigma_r = 0.5$ fm).
The longitudinal envelop functions $H^s_{P/T}$ and $H^n_{P/T}$ are parametrized as follows:
\begin{align}
H^s_{P/T}(\eta_s) &= \theta(\eta_{\text{max}} - |\eta_s|)\left(1\pm\frac{\eta_s}{y_{\text{beam}}}\right) \left[\theta(\eta_0^s - |\eta_s|)\right. \nonumber\\
&\left.+ \theta(|\eta_s|-\eta_0^s) \exp\left(\frac{\left(|\eta_s|-\eta_0^s\right)^2}{2\sigma_s^2}\right) \right]\,,
\\
H^n_{P/T}(\eta_s) &= \frac{1}{N} \left[ \theta(\eta_s-|\eta^n_{0;P/T}|)\exp\left(- \frac{(\eta_s-\eta^n_{0;P/T})^2}{2\sigma_{n;P/T}^2} \right) \right. \notag \\
&+\left. \theta(|\eta^n_{0;P/T}|-\eta_s)\exp\left(- \frac{(\eta_s-\eta^n_{0;P/T})^2}{2\sigma_{n;T/P}^2} \right)  \right]\,.
\end{align}
The parameters in the above envelope functions are fixed by comparing to the final rapidity distributions of charged hadrons. Table \ref{table:parameter} shows all the parameters used in the Monte-Carlo Glauber model for initial conditions in Au+Au collisions at different collision energies. Note that in this work, for each collision energy the centrality is classified according to the impact parameter. For each centrality class, we perform 500 event-by-event hydrodynamic simulations. {  In this work, we set the initial flows in the transverse and space-time rapidity directions to be zeros, i.e., $U_x=U_y =U_{\eta_s}=0$. }

\begin{table}[h]
\centering
\vline
\begin{tabular}{c|c|c|c|c|c|c|c|}
\hline
$\sqrt{s_{NN}}$ [GeV]&K & $\tau_0$ [fm] & $\sigma_s$ [fm] & $\eta_0^s$ & $\sigma_{n;P}$ & $\sigma_{n;T}$ &$\eta_{0;P/T}^n$  \\ \hline
7.7  & 7.67  & 3.6 & 0.3 & 0.9  & 0.07 & 0.7  &1.05\\ \hline
14.5 & 9.22  & 2.2 & 0.3 & 1.15 & 0.14 & 0.81 &1.4 \\ \hline
19.6 & 10.22 & 1.8 & 0.3 & 1.3  & 0.14 & 0.85  &1.5 \\ \hline
27   & 10.35 & 1.4 & 0.3 & 1.6  & 0.14 & 1.06  &1.8 \\ \hline
39   & 10.35 & 1.3 & 0.3 & 1.9  & 0.14 & 1.13  &2.2 \\ \hline
62.4 & 10.8  & 1.0 & 0.3 & 2.25 & 0.14 & 1.34  &2.7 \\ \hline
\end{tabular}
\caption{The parameters for a 3-dimensional Monte-Carlo Glauber model for initial conditions.}
\label{table:parameter}
\end{table}

\subsection{Viscous hydrodynamics}

To take into account the effect of finite net baryon density at RHIC-BES energies, our integrated event-by-event CLVisc framework includes both energy-momentum and baryon number conservation:
\begin{align}
&\nabla_{\mu} T^{\mu\nu}=0 \, ,\\
&\nabla_{\mu} J^{\mu}=0  \, ,
\end{align}
where $\nabla_{\mu}$ represents the covariant derivative operator in the Melin coordinate.
The energy-momentum tensor $T^{\mu\nu}$ and net baryon current $J^{\mu}$ take the following form:
\begin{align}
&T^{\mu\nu} = eU^{\mu}U^{\nu} - P\Delta^{\mu\nu} + \pi^{\mu\nu}\,, \\	
&J^{\mu} = nU^{\mu}+V^{\mu}\,,
\end{align}
where $e$ is the energy density, $U^{\mu}$ is the flow velocity, $P$ is the pressure, $\Delta^{\mu\nu} = g^{\mu\nu} - U^{\mu}U^{\nu}$, $\pi^{\mu\nu}$ is the shear-stress tensor, $n$ is the net baryon density, and $V^{\mu}$ is baryon diffusion current.
{  In this work, we follow a few recent studies \citep{Shen:2017bsr, Akamatsu:2018olk, Shen:2020jwv, Denicol:2018wdp} and neglect the bulk viscosity in the evolution equations for the sake of simplicity and saving the computing time.
We would like to point out that the bulk viscosity does have important effects on the medium evolution, especially near the region of the phase transition.
It will typically reduce the radial flow of the bulk matter as shown by several previous studies at the LHC energies \citep{Ryu:2015vwa, Ryu:2017qzn}.
In order to compensate the effect of bulk viscosity on the radial flow, we have adjusted the initial proper time $\tau_0$ to reproduce the final charged particle mean transverse momenta $\langle p_T\rangle$.
The inclusion of the bulk viscosity in the evolution equations will be left for a future study.}
The dissipative currents $\pi^{\mu\nu}$ and $V^{\mu}$ are evolved according to the following Israel-Stewart-like equations \citep{Denicol:2018wdp}:
\begin{align}
\Delta^{\mu\nu}_{\alpha\beta}D\pi^{\alpha\beta} &= -\frac{1}{\tau_{\pi}}\left(\pi^{\mu\nu} - \eta_v\sigma^{\mu\nu}\right)
\\
&- \frac{4}{3}\pi^{\mu\nu}\theta-\frac{5}{7}\pi^{\alpha<\mu}\sigma_{\alpha}^{\nu>}
+ \frac{9}{70}\frac{4}{e+P}\pi^{<\mu}_{\alpha}\pi^{\nu>\alpha}\,,
\nonumber
\\
\Delta^{\mu\nu}DV_{\nu}  &=  - \frac{1}{\tau_V}\left(V^{\mu}-\kappa_B\bigtriangledown^{\mu}\frac{\mu_B}{T}\right)-V^{\mu}\theta-\frac{3}{10}V_{\nu}\sigma^{\mu\nu}\,,
\end{align}
where $\theta$ is the expansion rate, $\sigma^{\mu\nu}$ is the symmetric shear tensor, $\eta_v$ and $\kappa_B$ are the transport coefficients for the evolution of shear tensor and baryon diffusion current. In this work, we consider the specific shear viscosity $C_{\eta_v}$ and baryon diffusion coefficient $\kappa_B$ as model parameters, which are related to $\eta_v$ and $k_B$ as follows:
\begin{align}
C_{\eta_v} &= \frac{\eta_v T}{e+P}, \\
\kappa_B &= \frac{C_B}{T}n\left(\frac{1}{3} \cot \left(\frac{\mu_B}{T}\right)-\frac{nT}{e+P}\right) \,.
\end{align}
One can see that the specific shear viscosity $C_{\eta_v}$ reduces to $\eta_v/s$ for zero net baryon density $n=0$ (used at top RHIC and LHC energies). In this work, we set $C_{\eta_v} = 0.08$ and $C_B = 0.4$ as constants for all collision energies.
The relaxation times are chosen as $\tau_{\pi} = \frac{5C_{\eta_v}}{T}$ and $\tau_V = \frac{C_B}{T}$.

To solve the hydrodynamic conservation and dissipation equations numerically, the Kurganov-Tadmor (KT) algorithm \cite{Pang:2018zzo,Gale:2013da,KURGANOV2000241,Schenke:2010nt} with second-order Runge-Kutta method is utilized to obtain the dynamical evolution of local macroscopic thermodynamic quantities (energy density $e$, net baryon density $n$ and flow velocity $U^{\mu}$).
KT algorithm is a finite volume method with the advantages of small numerical viscosity and clear physical interpretation; the change of average conserved quantities in a cell is determined by the flux at the cell interfaces.
{  In order to keep numerical stabilities when solving the dissipative hydrodynamics equations, we compare, at each cell, the maximum components of shear tensor $\pi^{\mu\nu}$ and baryon diffusion current $q^{\mu}$ to the ideal parts of energy-momentum tensor and net baryon current. If ${\rm max}(|\pi^{\mu\nu}|) > T^{\tau\tau}_{\rm ideal}$ or ${\rm max}(|V^{\mu}|) > J^{\tau}_{\rm ideal}$, we locally set $\pi^{\mu\nu} =0 $ or $V^{\mu} =0$.  Since the cells that need regulations are usually located at the dilute region, such treatment should not affect the major part of medium evolution.}

To close the equations of motion for the relativistic hydrodynamics, equation of state of the QCD medium has to be supplied. Several models have implemented the first order phase transition and the critical point in equation of states \citep{Parotto:2018pwx}. In this work, we use the NEOS-BQS equation of state \citep{Monnai:2019hkn,Monnai:2021kgu}. NEOS-BQS is based on the lattice QCD simulation at high temperature and vanishing net baryon density, and utilizes Taylor expansion method to construct the equation of state at finite net baryon density. At lower energy density, it matches the hadron gas equation of state via a smooth crossover.

\subsection{Particlization and afterburner}

When the local energy density drops below the freeze-out energy density (we set $e_{\text{frz}}$= 0.4~GeV/fm$^3$), the Cooper-Frye formula is used to obtain the momentum distribution of particles:
\begin{align}
\frac{dN}{dY p_T dp_T d\phi} = \frac{g_i}{(2\pi)^3}\int_{\Sigma} p^{\mu}d\Sigma_{\mu}f_{eq}(1+\delta f_{\pi}+\delta f_{V})\,.
\end{align}
In the above equation, $g_i$ is the degeneracy for identified hadrons; $d\Sigma_{\mu}$ is the hyper-surface element which is determined from the Cornelius routine \cite{Huovinen:2012is}; $f_{\rm eq}$, $\delta f_{\pi}$ and $\delta f_{V}$ are thermal equilibrium distribution and out-of-equilibrium corrections, which take the following forms:
\begin{align}
	f_{\rm eq} &= \frac{1}{\exp \left[(p_{\mu}U^{\mu} - B\mu_B \right)/T_f] \pm 1} \, ,\\
	\delta f_{\pi}(x,p) &= (1\pm f^{\text{eq}}(x,p)) \frac{p_{\mu}p_{\nu}\pi^{\mu\nu}}{2T^2_f(e+P)}, \\
	\delta f_V(x,p) &= (1\pm f^{\text{eq}}(x,p))\left(\frac{n}{e+P}-\frac{B}{U^{\mu}p_{\mu}}\right)\frac{p^{\mu}V_{\mu}}{\kappa_B/ \tau_V },
\end{align}
where $T_f$ is the chemical freeze-out temperature, $\mu_B$ is the net baryon chemical potential, $B$ is the baryon number for the identified baryon, $n$ is the local net baryon density.
Note that the above forms of the out-of-equilibrium corrections $\delta f_{\pi}$ and $\delta f_{V}$ are derived from Boltzmann equation via relaxation time approximation \citep{McNelis:2021acu}.
Due to nonzero net baryon density, $T_f$ and $\mu_B$ are different for each hypersurface cell.
This will lead to different particle density when particles are sampled in the comoving frame of the fluid.
Note that the thermodynamic variables on the hypersurface are calculated in terms of SMASH hadron resonance gas EOS to be consistent with particle species in SMASH.
{   In this work, we follow a few recent studies~\citep{Pang:2018zzo,Karpenko:2015xea,Shen:2014vra,Shen:2014lye} and use the Monte-Carlo method to sample the positions and momenta of thermal hadrons according to the Cooper-Frye formula.
We use a step function $\theta(f_{\rm eq} +\delta f_{\pi}+\delta f_{V})$ so that the full (equilibrium + non-equilibrium) distribution functions never encounter negative values. Such a treatment should have small effect on the final results since hydrodynamics mainly focuses on the soft physics while large viscous corrections usually appear at large $p_T$ region.
Recently, the Pratt-Torrieri-Bernhard (PTM) distribution~\citep{McNelis:2021acu,Pratt:2010jt,Du:2019obx} and the maximum-entropy distribution~\citep{Everett:2021ulz} have been proposed to treat the large viscous corrections on the freeze-out hypersurface. In the future, we will implement the PTM distribution or maximum-entropy distribution in our CLVisc framework.}

After the particlization of the fluid, the evolution of hadrons is described by the microscopic transport model SMASH \citep{Weil:2016zrk,Schafer:2019edr,Mohs:2019iee,Hammelmann:2019vwd,Mohs:2020awg} which solves the relativistic Boltzmann equation with elastic collisions, resonance excitations, string excitations and decays for all mesons and baryons up to mass $\sim$ 2~GeV.
To improve the statistics, we repeat the sampling and SMASH simulation 2000 times for each single hydrodynamic event, then average over 2000 SMASH events for final analysis.

To reduce the large amount of computing cost for event-by-event hydrodynamic simulations on central processing units (CPUs), CLVisc framework is accelerated on GPUs using OpenCL languages.
For example, using the grid size $N_x \times N_y \times N_{\eta} = 201 \times 201 \times 201$, running a hydrodynamics event with a smooth Glauber initial condition for 0-5\% Au+Au collisions at $\sqrt{s_{NN}} = 19.6$~GeV on GeForce GTX 3080 GPU typically takes $\sim 1800$ seconds or $\sim 1.6$ seconds per time step.
The particle sample module takes  $\sim 0.066$ seconds per sample event, and SMASH takes $\sim 2.5$ seconds per afterburner event.
Parallelization on GPUs provides a remarkable improvement on computing time for (3+1)-dimensional viscous hydrodynamics simulation, which makes it possible to study the event-by-event properties of QGP at BES energies.

\begin{figure*}[tbh]
\includegraphics[width=0.495\textwidth]{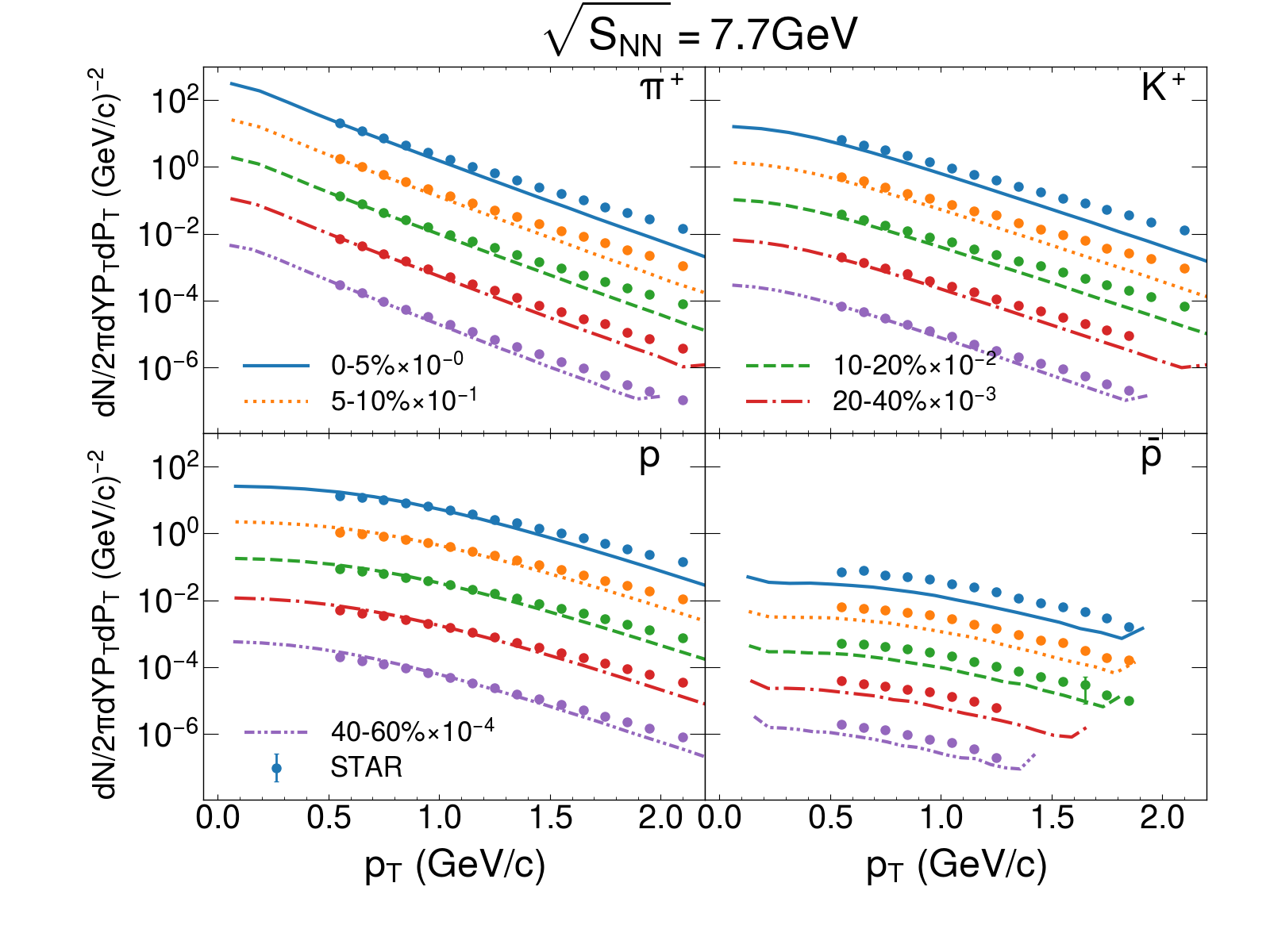}
\vspace{-4pt}
\includegraphics[width=0.495\textwidth]{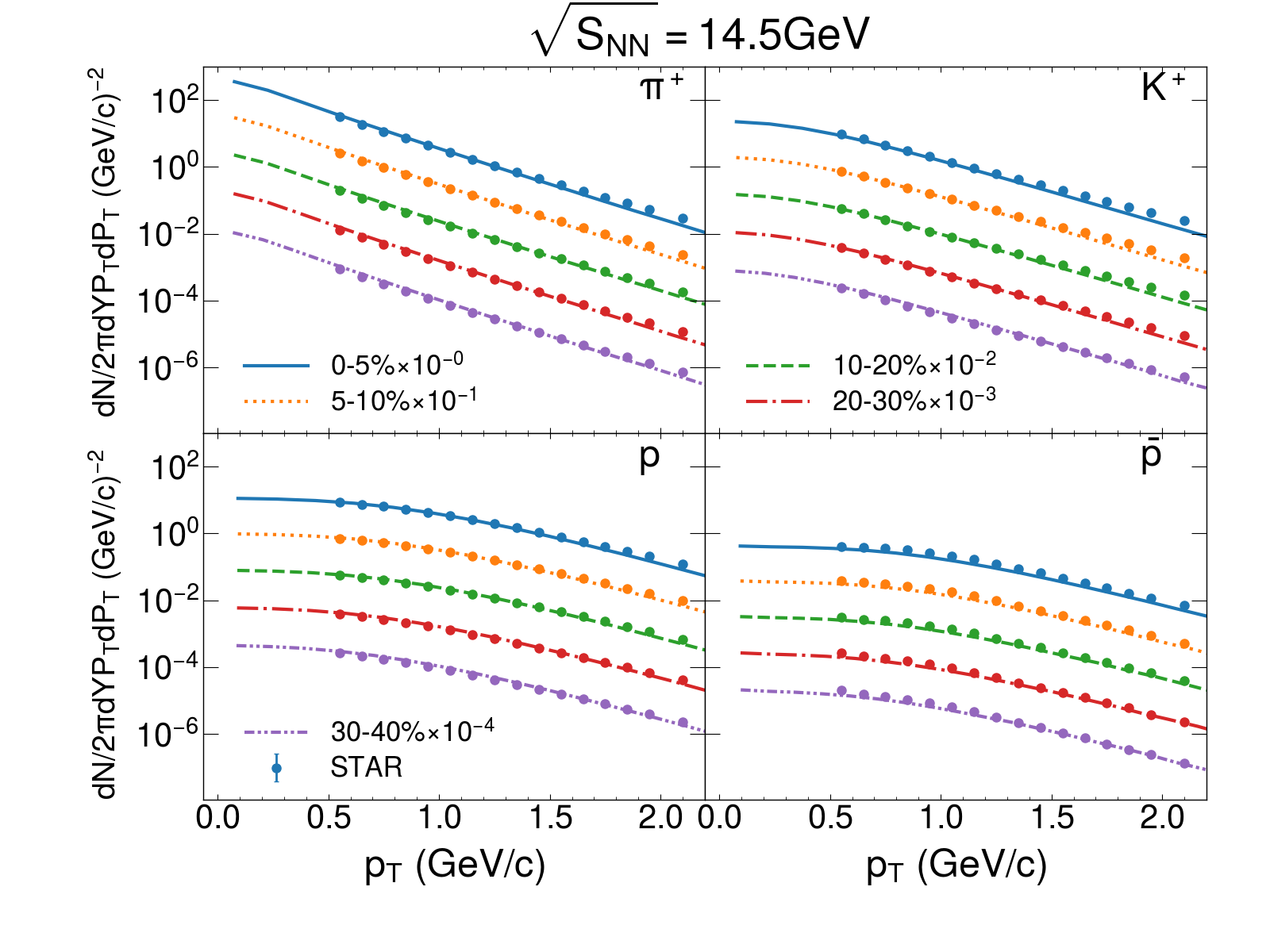}
\includegraphics[width=0.495\textwidth]{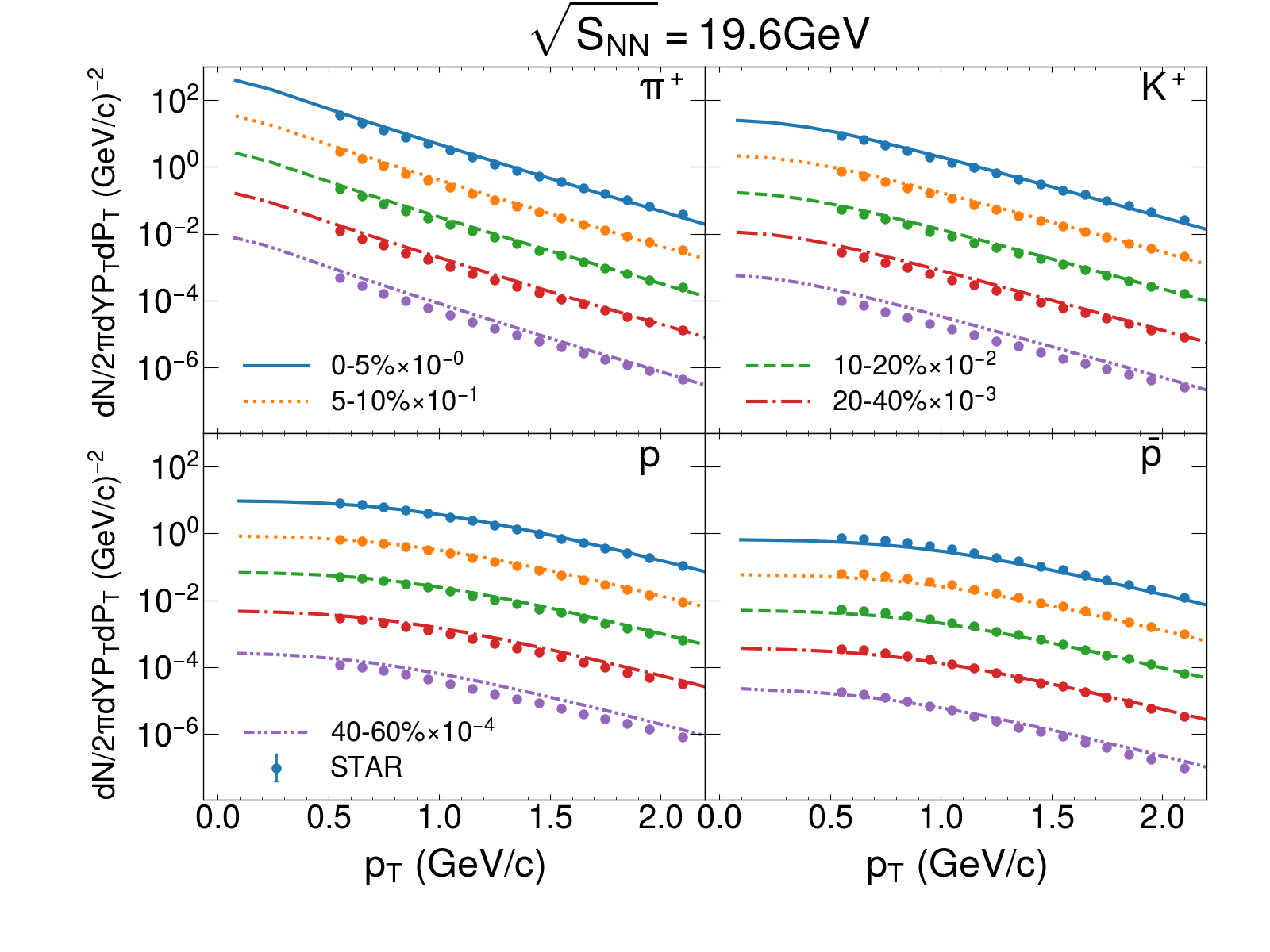}
\vspace{-4pt}
\includegraphics[width=0.495\textwidth]{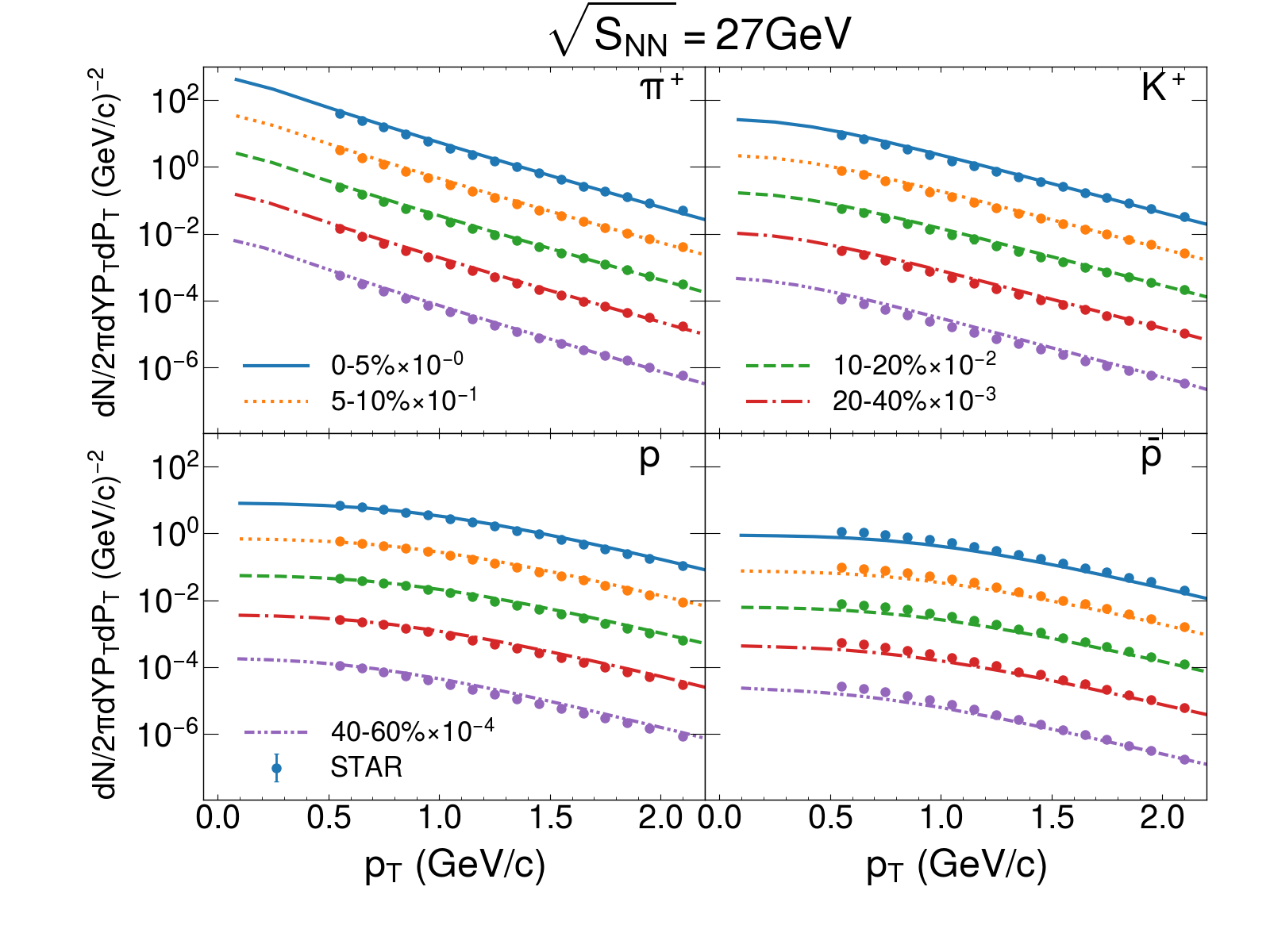}
\includegraphics[width=0.495\textwidth]{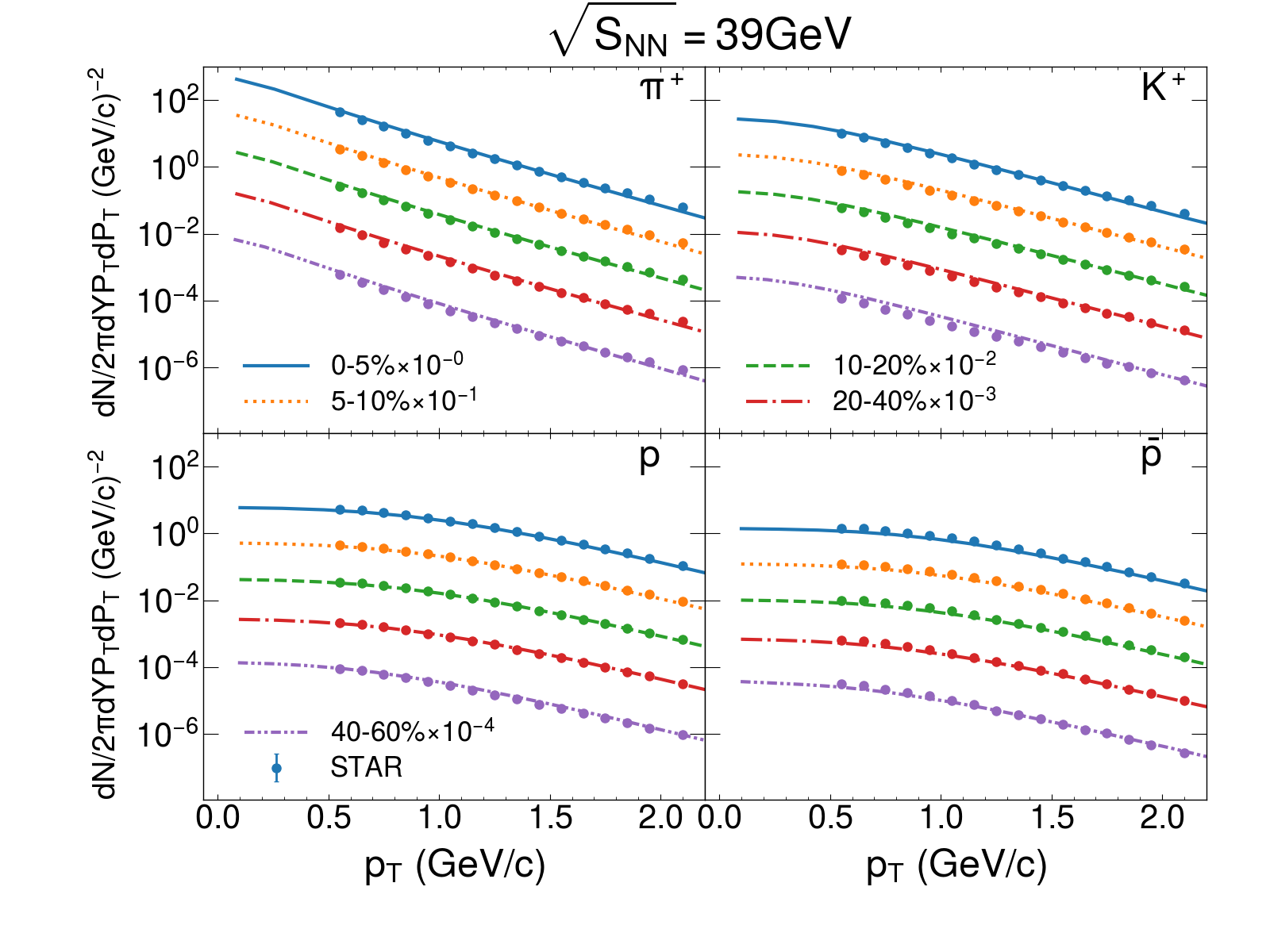}
\vspace{-4pt}
\includegraphics[width=0.495\textwidth]{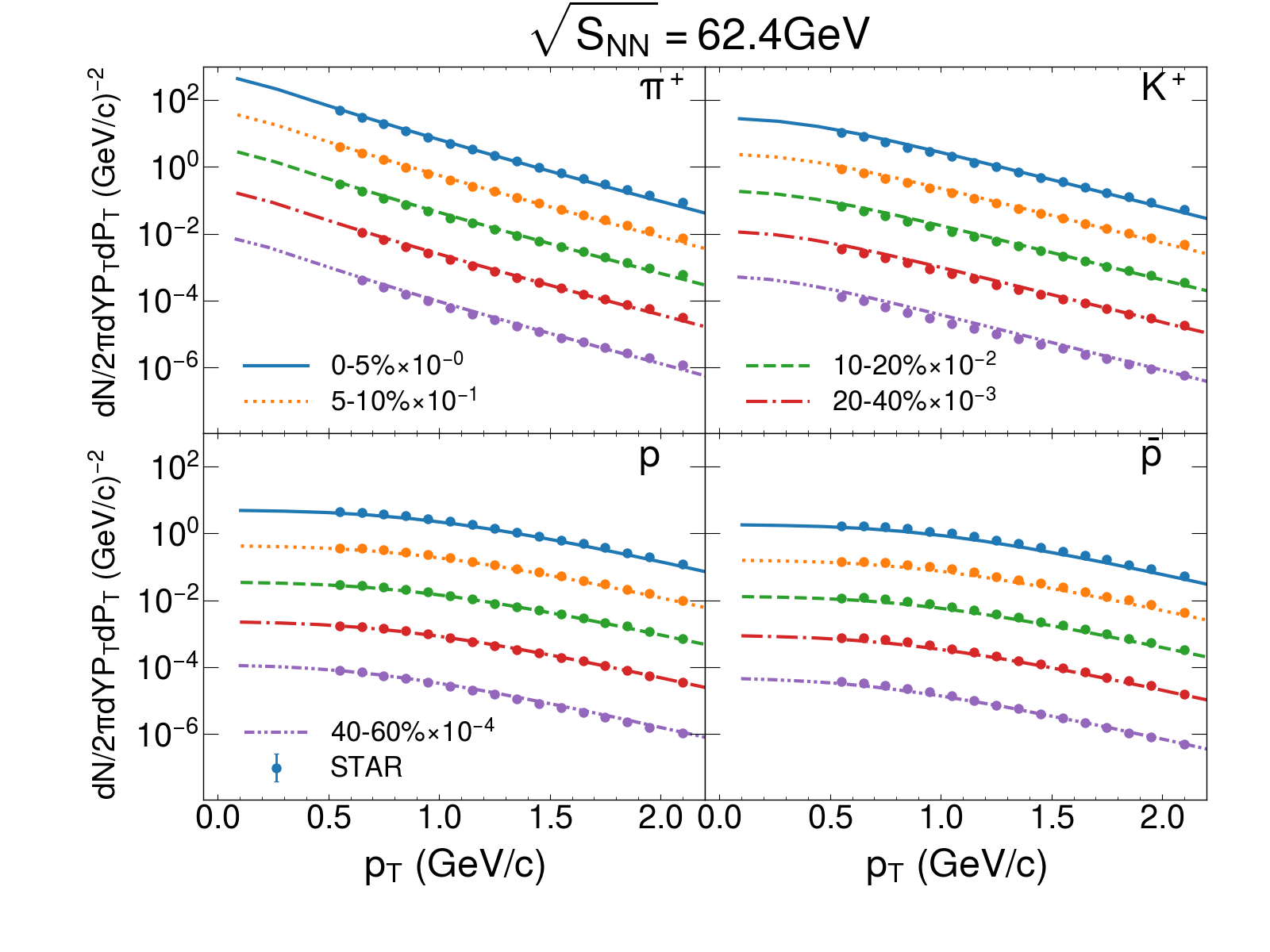}
\caption{The transverse momentum spectra for identified particles ($\pi^+$, $K^+$, p and $\bar{p}$) in different centrality classes in Au+Au collisions at $\sqrt{s_{NN}}$= 7.7, 14.5, 19.6, 27, 39 and 62.4~GeV. The data are taken from STAR collaboration \citep{Adamczyk:2017nof}.}
\label{particle_spectra}
\end{figure*}

\begin{figure*}[tbh]
\hspace{-24pt}
\includegraphics[width=0.995\textwidth]{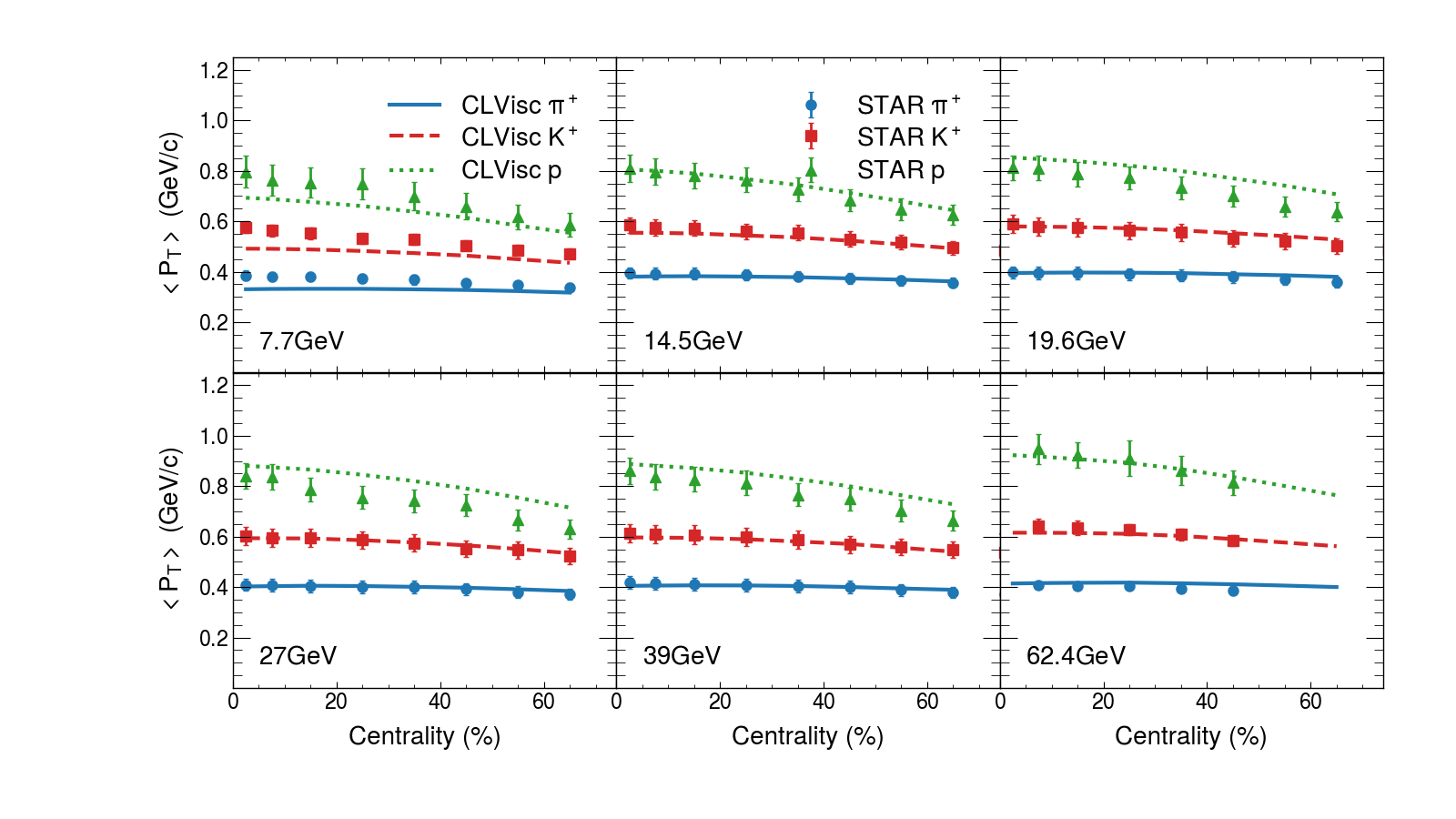}
\vspace{-12pt}
\caption{The centrality dependence of mean transverse momenta for $\pi^+$, $K^+$ and $p$ in Au+Au collisions at $\sqrt{s_{NN}}$ = 7.7, 14.5, 19.6, 27, 39, 62.4~GeV. The data are taken from STAR \citep{Adamczyk:2017iwn,Abelev:2008ab}.}
\label{mean_pt}
\end{figure*}

\begin{figure*}[tbh]
\includegraphics[width=0.495\textwidth]{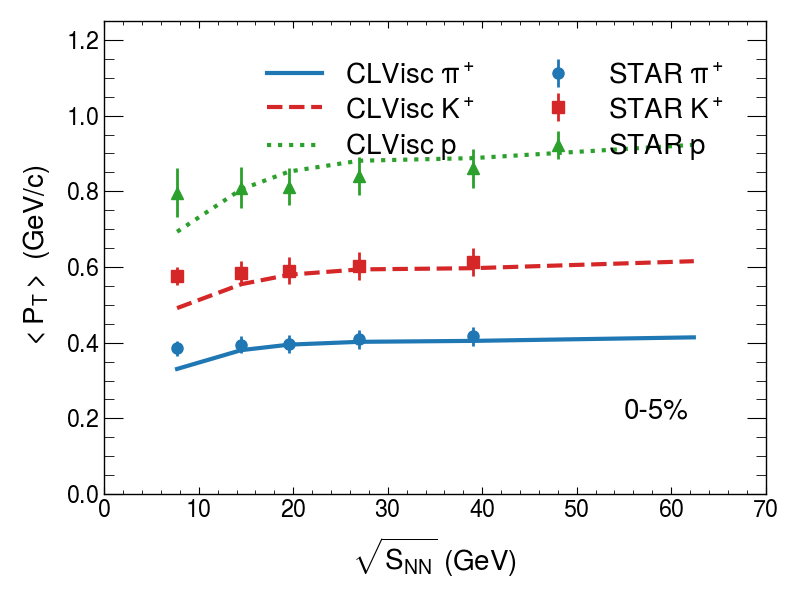}
\label{averagept05}
\includegraphics[width=0.495\textwidth]{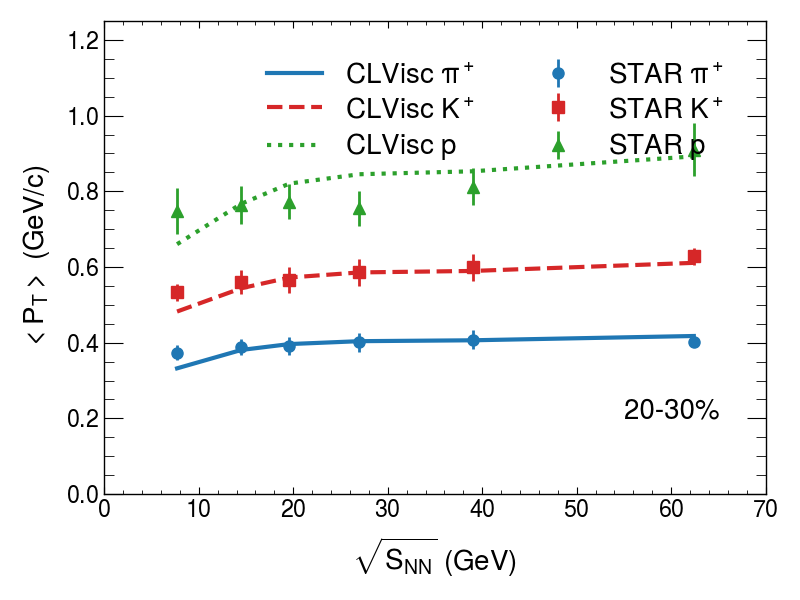}
\caption{ The collision energy dependence of mean transverse momentum for $\pi^+$, $K^+$ and $p$ in Au+Au collisions at 0-5\% and 20-30\% centrality. The data are taken from STAR \citep{Adamczyk:2017iwn,Abelev:2008ab}.}
\label{mean_pt_snn}
\end{figure*}

\begin{figure*}[tbh]
\hspace{-24pt}
\includegraphics[width=0.995\textwidth]{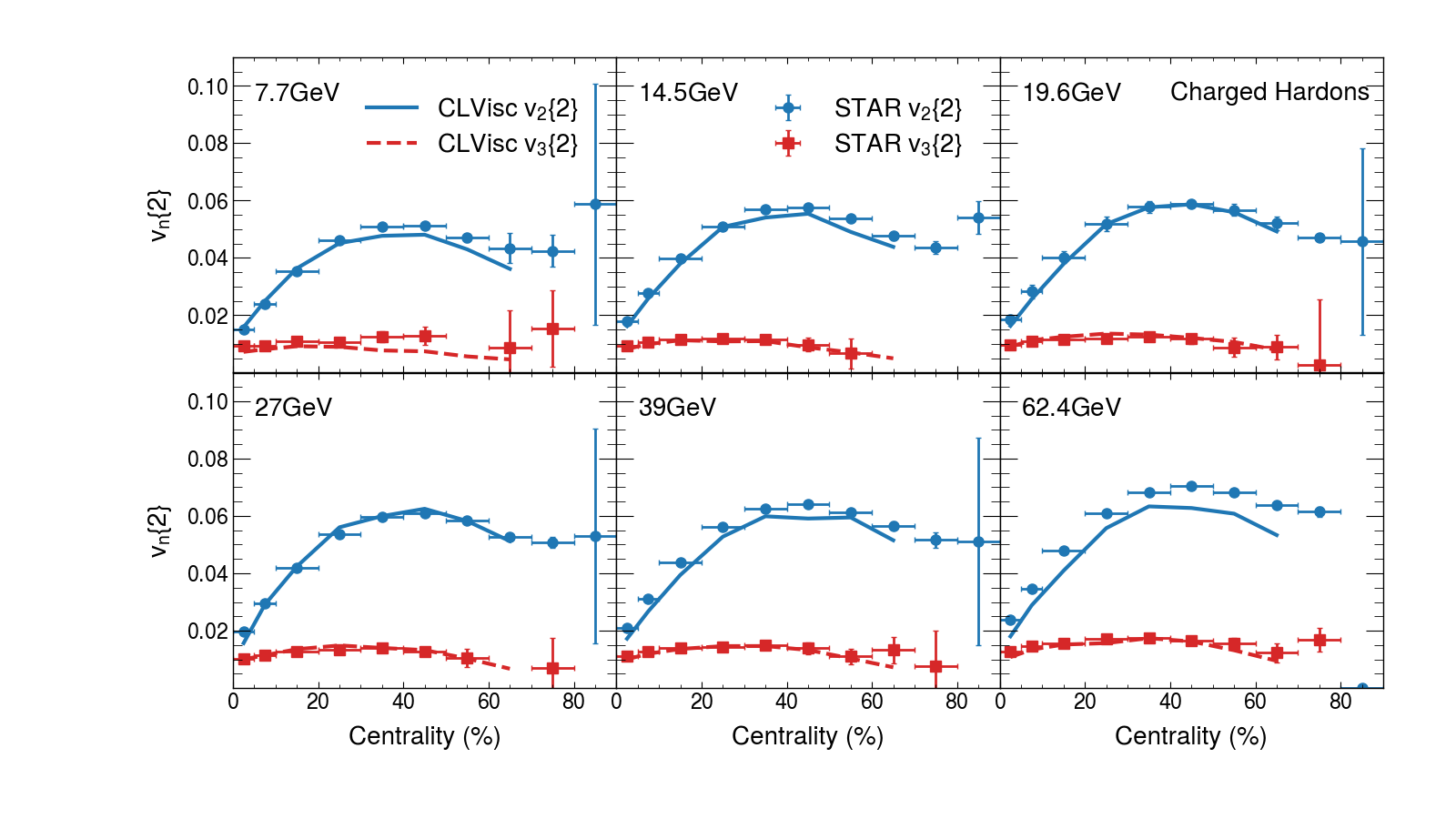}
\vspace{-12pt}
\caption{The centrality dependence of elliptic flow and triangle flow for charged hadrons within $|\eta|<1$ and $p_T > 0.2$ GeV in  AuAu collisions at $\sqrt{s_{NN}}$ = 7.7, 14.5, 19.6, 27, 39, 62.4 GeV. The data are taken from STAR \citep{Adamczyk:2017hdl,Adamczyk:2016exq}. }
\label{vn2_all}
\end{figure*}

\begin{figure*}[tbh]
\includegraphics[width=0.495\textwidth]{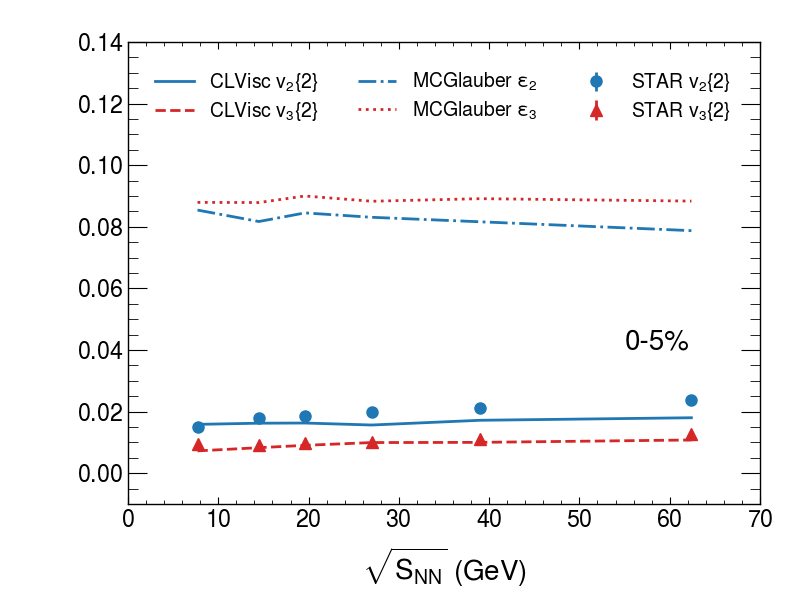}
\label{vn2_05}
\includegraphics[width=0.495\textwidth]{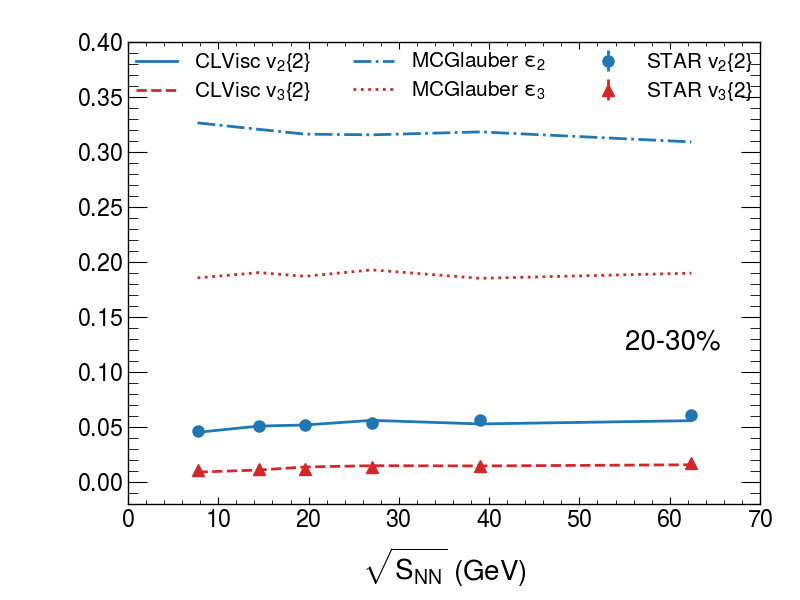}
\caption{The collision energy dependence of initial eccentricity, initial triangularity, final elliptic flow and final triangle flow for charged hadrons in  AuAu collisions at 0-5\% and 20-30\% centrality. The data are taken from STAR \citep{Adamczyk:2017hdl,Adamczyk:2016exq}. }
\label{vn2_all_snn}
\end{figure*}

\begin{figure}[tbh]
\includegraphics[width=0.495\textwidth]{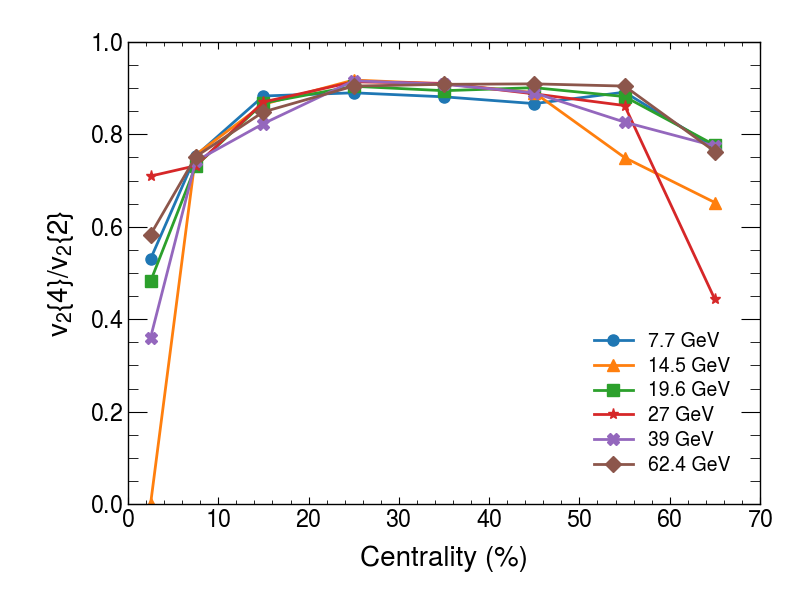}
\caption{The centrality dependence of multi-particle cumulants ratio $v_2\{4\}/v_2\{2\}$ for charged hadrons in  AuAu collisions at $\sqrt{s_{NN}}$ = 7.7, 14.5, 19.6, 27, 39, 62.4 GeV. }
\label{vn2_vn4}
\end{figure}

\section{Numerical Results}
\label{results}

In this section, we present the numerical results for the bulk observables in Au+Au collisions at RHIC-BES energies (from 7.7 to 62.4~GeV) using our event-by-event relativistic hydrodynamics model CLVisc with finite net baryon density.
Since there is no critical point in NEOS-BQS equation of state, our calculation provides a benchmark for understanding the RHIC-BES data and serves as baseline for studying the possible critical behaviors of QCD matter at BES energies.

\subsection{Identified particle spectra}

In Figure \ref{particle_spectra}, we show the transverse momentum ($p_T$) spectra for identified particles $\pi^+$, $K^+$, $p$ and $\bar{p}$ in Au+Au collisions at BES energies (7.7-62.4~GeV) with 5 different centrality classes.
One can see that our numerical results are in good agreement with STAR data \citep{Adamczyk:2017nof} on the $p_T$ spectra at mid-rapidity range $|Y|<0.25$ without the contribution from weak decays except at 7.7 GeV.
This means that our integrated event-by-event CLVisc hydrodynamics model with finite net baryon density can describe the bulk evolution and radial flow of the QCD matter created in heavy-ion collisions at BES energies.

In order to see the $p_T$ spectra and the radial flow effect more clearly, we can look at the mean transverse momenta of identified particles which strongly depend on the slope of the $p_T$ spectra.
Figure \ref{mean_pt} shows the mean $p_T$ as a function of the centrality class in Au+Au collisions at BES energies (7.7-62.4~GeV).
One can see that our integrated model can describe most of the experimental data from STAR \citep{Adamczyk:2017iwn,Abelev:2008ab} except the lowest collision energy (7.7~GeV) explored here.
For Au+Au collisions at $\sqrt{s_{NN}}=7.7$~GeV, the overlap time is quite large ($\sim 3$~fm) and the pre-equilibrium stage can produce some additional radial flow which is neglected in our model.
The use of dynamical initial conditions and pre-equilibrium evolution should improve the model in the future.

In Figure \ref{mean_pt_snn}, we show the collision energy dependence of the mean $p_T$ of identified particles.
One can clearly see different radial flow effects for $\pi^+$, $K^+$ and $p$ (due to different masses): more blue shift effect for more massive particles.
As for the collision energy dependence, we find that the mean momenta of $\pi^+$, $K^+$ and $p$ increase mildly with the collision energy due to larger radial flow.

\subsection{Anisotropic flows and flow fluctuations}

Anisotropic collective flows, which originate from initial geometric anisotropies and fluctuations, are very important observables for studying the transport properties of QGP produced in relativistic heavy-ion collisions. In this work, we use the $\boldsymbol{Q_n}$ cumulant method \citep{Bilandzic:2010jr} to compute the $n$-th order anisotropic flows. In this method, one first defines the $\boldsymbol{Q_n}$ vector:
\begin{align}
\boldsymbol{Q_n}=\sum_{i=1}^{N} e^{in\phi_i} \, ,
\end{align}
where $N$ is the charged multiplicity for each event, and $\phi_i$ is the azimuthal angle for the $i$-th particle in the final state. Then the single-event-averaged $2$-particle and $4$-particle correlations can be defined as follows:
\begin{align}
\left\langle 2 \right\rangle &= \frac{|Q_n|^2-N}{N(N-1)} \nonumber \, , \\
\left\langle 4 \right\rangle &= \frac{|Q_n|^4 + |Q_{2n}|^2 - 2\text{Re}[Q_{2n}(Q^*_n)^2]}{N(N-1)(N-2)(N-3)} \nonumber \\
&- 2 \frac{2(N-2)|Q_n|^2 -N(N-3)}{N(N-1)(N-2)(N-3)}\, .
\end{align}
In the above flow correlations  $\left\langle 2 \right\rangle$ and $\left\langle 4 \right\rangle$, the self-correlations and finite particle number effects have been subtracted.
Then the 2-particle and 4-particle cumulants can be obtained from $\langle 2\rangle$ and $\langle 4\rangle$ by averaging over many events:
\begin{align}
C_n\{2\} &= \left\langle \left\langle 2 \right\rangle \right\rangle \,, \\
C_n\{4\} &= \left\langle \left\langle 4 \right\rangle \right\rangle - 2\left\langle \left\langle 2 \right\rangle \right\rangle ^2 \,.
\end{align}
Finally, the integrated anisotropic flow coefficients can be calculated as:
\begin{align}
v_n\{2\} &= \sqrt{C_n\{2\}} \,, \\
v_n\{4\} &= (-C_n\{4\})^\frac{1}{4} \,.
\end{align}
Figure \ref{vn2_all} shows the centrality dependence of elliptic flow $v_2\{2\}$ and triangle flow $v_3\{2\}$ for charged hadrons in Au+Au collisions at 7.7-62.4~GeV using two-particle cumulant method.
Here we only consider particles with $p_T > 0.2$ GeV and $|\eta|<1$ in the flow analysis.
One can see that our results are in good agreement with the experimental data from STAR \citep{Adamczyk:2017hdl,Adamczyk:2016exq}.
For the elliptic flow, one can clearly see the typical non-monotonic centrality dependence which originates from the combined effect of the elliptic geometry, geometrical fluctuations and the sizes of the collision systems.
From central to mid-central to peripheral collisions, the elliptic flow first increases from a finite value in the most central collisions caused by the geometrical fluctuations to the maximum value due to the increased eccentricity, and then decreases due to smaller system sizes.
The triangular flow has much weaker dependence on the collision centrality because it mainly originates from the initial state geometrical fluctuations.

In Figure \ref{vn2_all_snn}, we show elliptic flow $v_2\{2\}$ and triangular flow $v_3\{2\}$ in most central (0-5\%) and mid-central (20-30\%) Au+Au collisions as a function of the collision energy (7-62.4~GeV).
One can see that for the most central 0-5\% collisions, our results on elliptic flow  underestimate the experimental data a little bit.
One possible reason is the use of constant shear viscosity in our current study.
The inclusion of the temperature and baryon chemical potential dependences of the shear viscosity may improve the current result \citep{Shen:2020jwv}. This will be left for a future study.
As for the collision energy dependence, we find that both elliptic and triangular flows increase slightly with the beam energy.
This is mainly from the increase of radial flow due to the increase of initial energy density since the eccentricities $\epsilon_2$ and $\epsilon_3$ have very weak dependence on collision energy as shown in the figure.

The fluctuations of anisotropic flows are also valuable tools to probe the initial state fluctuations and the transport properties of the QGP.
In Figure \ref{vn2_vn4}, we show the multi-particle cumulant ratio $v_2\{4\}/v_2\{2\}$ as a function of centrality for different collision energies (7-62.4~GeV).
The deviation of this cumulant ratio from unity quantifies the relative flow fluctuations.
One can clearly see the non-monotonic dependence of the cumulant ratio $v_2\{4\}/v_2\{2\}$  on the collision centrality.
As one goes from central to mid-central to peripheral collisions, the multi-particle cumulant ratio $v_2\{4\}/v_2\{2\}$ first increases, and then decreases.
This means that the relative fluctuations are smallest in mid-central collisions, and are larger in central and peripheral collisions.
This is because the initial collision geometry dominates elliptic flow in mid-central collisions, while the fluctuations dominate elliptic flow in central and peripheral collisions. Another interesting observation is that the multi-particle cumulant ratio $v_2\{4\}/v_2\{2\}$ has weak collision energy dependence, consistent with the preliminary STAR data \citep{Magdy:2018itt,Niseemtalk}.

\section{Summary}
\label{summary}

In this work, we have developed an integrated event-by-event relativistic hydrodynamics framework to study the bulk properties of QCD matter at the finite net baryon density at RHIC-BES energies.
In particular, we have extended the CLVisc (3+1)-dimensional viscous hydrodynamics model to include the net baryon number conservation and the 2nd order Israel-Stewart-like equations for baryon diffusion current.
The NEOS-BQS equation of state with finite baryon chemical potential is utilized to close the equations of motion for hydrodynamics evolution.
The initial fluctuating energy density and net baryon density are constructed from Monte-Carlo Glauber model.
When the density of QCD matter drops below the chemical freeze-out, the dynamical evolution of the dilute hadron gas is simulated via the microscopic transport model SMASH.

Based on our integrated CLVisc hydrodynamics framework, we have first investigated identified particle spectra at RHIC-BES energies from 7.7 to 62.4~GeV.
Our model can reproduce the mass, centrality and collision energy dependences of the particle spectra and mean transverse momenta of identified particles.
We find that the mean transverse momenta increase (mildly) from peripheral to central collisions and from low to high beam energies mainly due to larger radial flow developed in larger and denser collision systems.
It is also interesting that the mean transverse momentum of protons is more sensitive to collision energies than those of $\pi^+$ and $K^+$, consistent with picture driven by the radial flow.

We have further computed the collision energy dependences of the anisotropic flows and flow fluctuations (in terms of the multi-particle cumulant ratio $v_2\{2\}/v_2\{4\}$) for charged hadrons.
We find that both elliptic and triangle flows increase mildly with the increase of collision energy.
Since the initial eccentricities have very weak dependence on collision energy, the mild increases of $v_2$ and $v_3$ mainly come from the increase of the radial flow when going to from low to high beam energies.
It is also interesting that the multi-particle cumulant ratio $v_2\{2\}/v_2\{4\}$ shows a non-monotonic dependence on the collision centrality; the relative fluctuations for $v_2$ are smallest in mid-central collisions.
This can be understood because the collision geometry dominates the elliptic flow in mid-central collisions.
In central and peripheral collisions the relative fluctuations of $v_2$ become larger since the elliptic flow is more dominated by fluctuations.
Our results also show that the relative fluctuations of $v_2$ are insensitive to collision energies, consistent with the preliminary STAR data.

To conclude, our new integrated event-by-event (3+1)-dimensional relativistic hydrodynamics CLVisc model provides a baseline framework for simulating the collective evolution of QCD matter at finite baryon density region, which is crucial for understanding the bulk properties of the QGP at RHIC-BES energies and for studying the critical properties of hot and dense QCD matter.
In the future, one may use this integrated model to test the equation of state with finite net baryon density \citep{Monnai:2019hkn,Monnai:2021kgu,Parotto:2018pwx}.
Our realistic hydrodynamics simulation can also provide the temperature and flow (gradient) profiles of the QGP, which are important inputs for studying jet quenching \cite{Qin:2015srf,Cao:2020wlm,Xing:2019xae,Chen:2020tbl,Li:2020kax,Cao:2016gvr} and the global/local polarizations \citep{Fu:2021pok,Pang:2016igs,Ryu:2021lnx,Becattini:2017gcx,Karpenko:2021wdm,Fu:2020oxj,Yi:2021ryh,Hidaka:2017auj,Karpenko:2016jyx,Fang:2016vpj,Becattini:2021iol} in relativistic heavy-ion collisions.
To further improve our model, we may use dynamical initial conditions \citep{Shen:2017bsr,Okai:2017ofp} and include pre-equilibrium evolution \citep{Kurkela:2018vqr,Kurkela:2018wud} in our event-by-event CLVisc hydrodynamics framework.
The extension of our CLVisc framework to include hydrodynamics fluctuations \cite{Young:2014pka, Sakai:2020pjw} is another interesting direction.
Studies along these directions can be pursued in the future.

\vspace{12pt}
\section{ACKNOWLEDGMENTS}

XYW would like to thank H.~Elfner for discussion and hospitality during his visit to Goethe University at Frankfurt.
This work is supported in part by Natural Science Foundation of China (NSFC) under Grants No. 11775095, No. 11890710, No. 11890711, No. 11935007, No. 12075098, No. 11221504, No. 11861131009 and No. 11890714, by U.S. Department of Energy under Grant No. DE-AC0205CH11231, and by U.S. National Science Foundation under Grants No. ACI-1550228 and No. OAC-2004571.
Some of the calculations were performed in the Nuclear Science Computing Center at Central China Normal University (NSC$^3$), Wuhan, Hubei, China.

\bibliographystyle{h-physrev5} %reference style
\bibliography{refs}   % data for reference

\end{document}